\documentclass{article} 
\usepackage{iclr2020_conference,times}

\usepackage{amsmath,amsfonts,bm}

\def\eqref#1{equation~\ref{#1}}

\def\1{\bm{1}}

\DeclareMathAlphabet{\mathsfit}{\encodingdefault}{\sfdefault}{m}{sl}
\SetMathAlphabet{\mathsfit}{bold}{\encodingdefault}{\sfdefault}{bx}{n}

\usepackage{hyperref}
\usepackage{url}

\usepackage{amsmath,amssymb,amsfonts}
\usepackage{algorithmic}
\usepackage{graphicx}
\usepackage{textcomp}
\usepackage{xcolor}
\usepackage[boxruled,linesnumbered]{algorithm2e}
\usepackage{graphicx} 
\usepackage{hyperref}
\usepackage[british]{babel}
\usepackage{hhline}
\usepackage{multirow}

\title{Topological based classification using graph convolutional networks}

\author{Roy Abel \\
Bar Ilan University\\
\texttt{royabel10@gmail.com} \\
\And
Idan Benami \\
Bar Ilan University\\
\texttt{idanbena@gmail.com} \\
\AND
Yoram Louzoun \\
Bar Ilan University\\
\texttt{louzouy@math.biu.ac.il}
}

\iclrfinalcopy 
\begin{document}

\maketitle

\begin{abstract}
In colored graphs, node classes are often associated with either their neighbors class or with information not incorporated in the graph associated with each node. We here propose that node classes are also associated with topological features of the nodes. We use this association to improve Graph machine learning in general and specifically, Graph Convolutional Networks (GCN). 

First, we show that even in the absence of any external information on nodes, a good accuracy can be obtained on the prediction of the node class using either topological features, or using the neighbors class as an input to a GCN. This accuracy is slightly less than the one that can be obtained using content based GCN.

Secondly, we show that explicitly adding the topology as an input to the GCN does not improve the accuracy when combined with external information on nodes. However,  adding an additional adjacency matrix with edges between distant nodes with similar topology to the GCN does significantly improve its accuracy, leading to results better than all state of the art methods in multiple datasets.

\end{abstract}

\section{Introduction and Related work}
One of the central assumptions in node classification tasks is that neighboring nodes have similar classes \citep{ji2012variance, berberidis2018data, zhu2003combining, sindhwani2005beyond}. This has been extensively used in node classification tasks \citep{belkin2004semi, zhu2003semi}. Such approaches are now often denoted as graph neural networks (i.e. machine learning where the input is a graph/network) \citep{scarselli2008graph, gori2005new, li2015gated}. Four main approaches have been proposed to take advantage of a graph in machine learning:
\begin{itemize}
\item Regularize the output requiring that neighboring nodes either in the graph or in its projection have similar classes.
\item Use the graph to propagate labels and learn the best propagation.
\item Use the graph to project the nodes to real valued vectors and use those for supervised or unsupervised learning.
\item Use Graph Convolutional Networks (GCN) for convolutions on the input of a node and its neighbors.
\end{itemize}

Regularization or graph partitioning methods include among others partitioning the graphs based on the eigenvalues of the Laplacian  (assuming that nodes with the same partition have similar classes). The Laplacian of a graph is : $L=D-A$  , where  $D$ is a diagonal matrix, with the sum of each row in the diagonal and $A$ is the adjacency matrix. This Laplacian is often weighted by multiplying it on the left and the right by $D$ to normalize for the degree \citep{dhillon2007weighted, karypis1995metis}. Other works have used variants of this idea, each using smoothness and graph distance differently \citep{belkin2004semi, sindhwani2005beyond}. An alternative approach is to use quadratic penalty with fixed labels for seed nodes \citep{zhou2004learning, zhu2003semi}.  

Multiple diffusion and information propagation models have  also been proposed either through explicit diffusion, or through the projection of nodes into real valued vectors \citep{rosenfeld2017semi}. For example, DeepWalk \citep{perozzi2014deepwalk}, where a truncated random walk is performed on nodes. It then uses these sentences as an input to skipgram to compute a projection of each word into $R^N$, maximizing the sentence probability.  Planetoid \cite{yang2016revisiting} also uses random walks combined with negative sampling. \citet{duvenaud2015convolutional} uses a translation of subgraphs to hash functions for a similar task in the context of molecule classifications. A very similar approach was presented by \citet{grover2016node2vec} (Node2Vec) by projecting nodes minimizing the distance of neighbored nodes in a truncated random walk. The DNGR model \citep{cao2016deep} uses random walk to compute the mutual information between points (the PPMI-positive pointwise mutual information), and then a SVD decomposition to project into space. PPMI was used for word representations in \citet{levy2015improving} and is a sparse high dimensional representation. 

Another possible approach is the projection of the graph (often using the Laplacian eigenvectors), and the usage of the projection for classification (and not only for a smoothness based regularization), where either the graph itself is used (in such a case, the eigenvectors themselves are used) or an input to the graph is used. In such a case, a convolution with these eigenvectors was used \citep{masci2015shapenet, monti2017geometric}. A Multi-Dimensional-Scaling (MDS) projection of the points in the graphs was also used for a similar goal \citep{belkin2002laplacian, levy2015improving}.   Alternative approaches were  inspired again by word embedding methods \citep{mikolov2013distributed} such as word2vec. These methods use the graph to define a “context” in relation to which the node embedding is constructed. When the data includes only the graph, the embeddings are used as features and fed into existing predictors \citep{perozzi2014deepwalk}. These methods can be thought of as propagating features rather than labels. \citet{henderson2011s} defines local features to translate each node to a features vector and use those to predict classes. 

Recently, Kipfs and collaborators, in a seminal work, proposed a simplification of spectral based convolutions \citep{kipf2016semi, schlichtkrull2018modeling}, and instead use a two-layer approach, which can be summarized as: 
\begin{equation} \label{GCN Layer}
X_{n+1}=\sigma(\tilde{A} \times X_n \times W_n ),
\end{equation}
 where $\tilde{A}$  is a normalized adjacency matrix: $\tilde{A}=D^{-1/2}[A+A^T+I]D^{-1/2}$. They test their work on multiple graphs with labeled nodes including CiteSeer, Cora, Pubmed, and Nell.  
 
Convolution approaches can also be used with the graph as a filter on the input. Most such convolutions are spectral (use the Laplacian eigenvectors). However, recent methods are based on random filters. Those include among others:  \citet{atwood2016diffusion} which defines predetermined convolutions with powers of the adjacency matrix and then combines these powers using learned weights to maximize the classification precision of either the full graph or the classification of nodes. \citet{bruna2013spectral} provide a multi-level graph convolution with pooling, where at each stage nodes are merged into clusters using agglomerative clustering methods, and combine it with a pooling method to represent the different resolution of images. This has been extended \citep{henaff2015deep, bronstein2017geometric} to different convolutional kernels (mainly spectral, but also diffusion-based kernels) and the classification of images, using ImageNet (see \citet{bronstein2017geometric} for a detailed review of all convolution methods).  Vandergheynst and collaborators mainly use polynomial convolution in the spectral domain. Similar formalisms were used to study not only single snapshots, but also with recurrent networks time series of graphs, mainly again in image analysis \citep{seo2018structured}.	 Over the last 3 years, over 1,500 extensions and applications of GCN have been published in combination with many other learning methods,  including among many others combinations of GCN with recurrent neural networks \citep{ling2019fast}, with GANs \citep{lei2019gcn} and with active learning \citep{abel2019regional}. 

GCNs capture dependencies of nodes' features. However, current techniques consider only local neighborhoods. Thus, long-range dependencies can only be captured when these operations are applied repeatedly, propagating signals progressively through the data. To catch long-range dependencies, \citet{kipf2016semi} proposed stacking multiple layers of GCN. While this is possible in theory, it has never been successfully applied. In practice, GCN models work the best with 2-3 layers \citep{kipf2016semi, monti2017geometric, velivckovic2017graph, levie2018cayleynets, fey2018splinecnn}. \citet{abu2018n} proposed using NGCN – train multiple instances of GCNs over different distances regions. While this led to good performance, it is highly inefficient and does not scale to long distances (as the number of models scales linearly with the desired length).

However, long range correlations can be obtained from a different direction. Recently, a correlation has been shown between the topological attributes (e.g. degree, centrality, clustering coefficient...) of nodes and their class \citep{shi2000normalized, yang2013community, cannistraci2013minimum, rosen2015topological, naaman2018edge}. Inspired by the improvement of non-local operations in a variety of tasks in the field of computer vision \cite{wang2018non}, we propose a novel non-local operation for GCN, based on the topology of the graph. Our operation is generic and can be implemented with every GCN to capture long-range dependencies, allowing information propagation to distant nodes. 

There are several advantages of using non-local operations: (a) In contrast to the standard local convolution layer, non-local operations capture long-range dependencies directly by computing interactions between any two nodes, regardless of their positional distance; (b) As we show in experiments, non-local operations are efficient and achieve their best results even with only a few layers; (c) Finally, our non-local convolution can be easily combined with other graph convolution techniques (e.g. GCN, GAT).

\section{Main Contributions Of The Current Work}
We here propose the following contributions of nodes topology to graph-based machine learning. 

First, we show that in the absence of external information, node topology can be used to predict the class of nodes using a feed-forward network.  The topology of a node is represented by a vector of attributes of each node, including among others, its degree, the frequency of different sub-graphs around it and its centrality.

We then show that this can be translated to GCN through an input representing the number of first and second neighbors belonging to each class in the training set. 

Finally, we show in the context of GCN, that it is better to add an additional adjacency matrix representing the similarity between node topologies to the GCN than actually adding the topology of the nodes as an input. 

GCN and Graph Attention Networks (GAT) with this additional adjacency matrix produce accuracies better than all state of the art methods on the Cora, Pubmed, and CiteSeer Datasets.

\section{Models And Data}
\subsection{Datasets studied} 
Following \citet{shchur2018pitfalls}, we used four well-known citation network datasets: PubMed, CiteSeer and CORA \citep{yang2016revisiting}, as well as the extended version of CORA from \citet{bojchevski2017deep}, denoted as CORA-Full, and two co-authorship networks: Coauthor CS, Coauthor Physics. Descriptions of these datasets, as well as statistics, can be found in Appendix~\ref{Appendix data sets}.

\subsection{Network structure}
We used the standard GCN model developed by \citet{kipf2016semi} or the GAT \citep{velivckovic2017graph}. 

Each GCN layer is defined as in Eq. 1, where $\tilde{A}$ is defined above,  $X_n$  is the input from the previous layer, and $W_n$ are the weights of the current layer.  

In GAT, each layer may contain multiple heads. A GAT head is a linear combination of
nodes' features followed by a non linear function:
\begin{equation} \label{GAT Layer}
h_i' = \sigma(\sum_{j \in N_i}\alpha_{i,j}Wh_j) 
\end{equation}
where $h_j$ is a set of features for node $j$, $W$ is a weight matrix, $\sigma$ is a non linear function, and $\alpha_{i,j}$ are the normalized attention coefficients. Attention coefficients are calculated for each pair of connected nodes to be:
\begin{equation}
\alpha_{i,j} = a(W  h_i, W  h_j)
\end{equation}
where $a$ is a single layer feed forward network, and $W$ is weight matrix.

\begin{figure} [t]
\begin{center}
 \includegraphics[width=10cm, height=6cm]{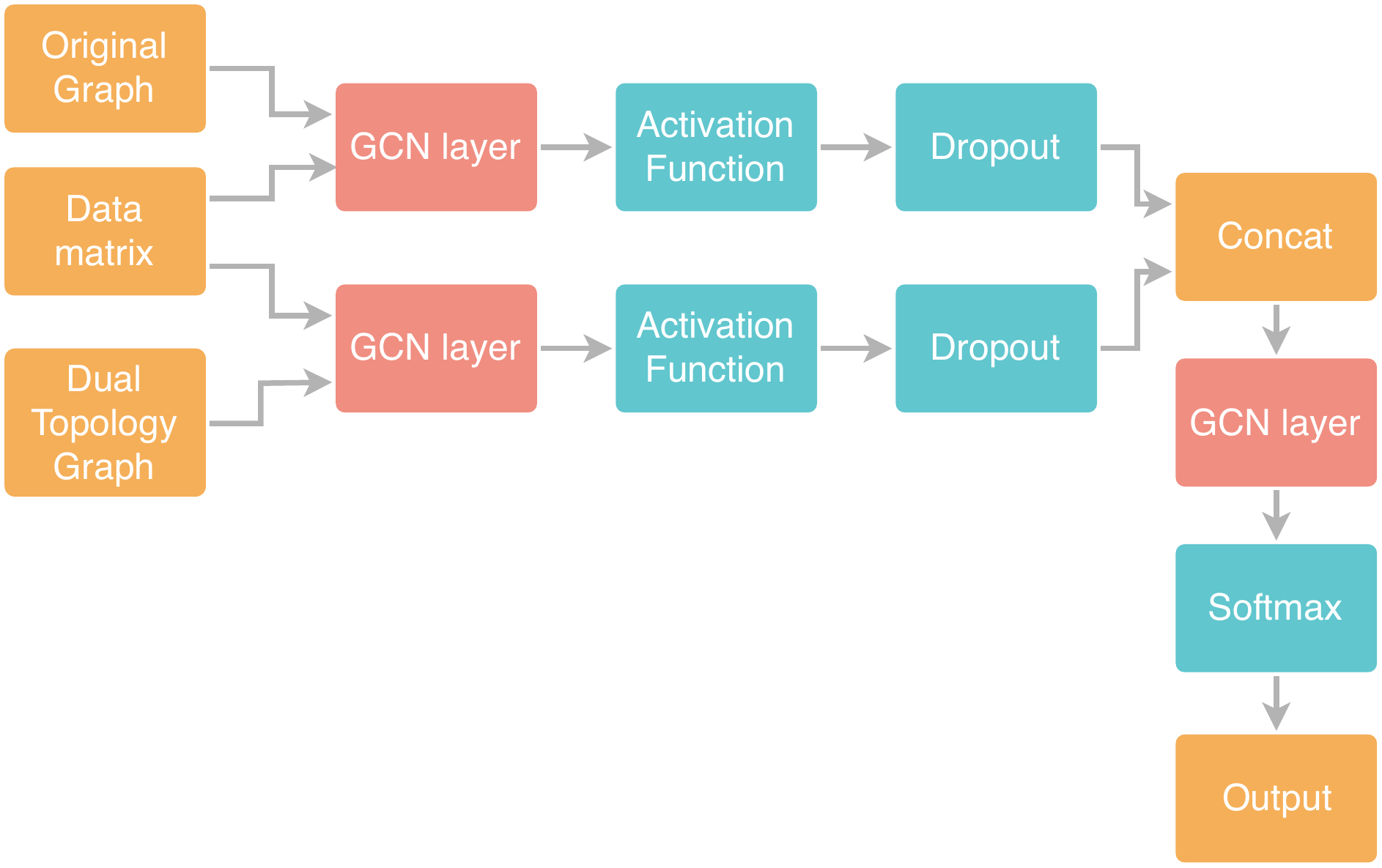}
\end{center}
\caption{\it{Schematic depiction of 2-layer Topology based Graph Convolutional Network (T-GCN). The first layer includes 2 parallel layers of GCN: the first on the original graph, and the second on the dual graph. The two outputs are then concatenated and used as an input to the second layer. The second layer is a standard GCN on the original graph}}

\label{Schematic T-GCN}
\end{figure}

The extensions we propose to the model come from either changing the input of the model or from altering $\tilde{A}$.  The following modifications were considered:
\begin{itemize}
\item Topology based GCN (T-GCN). We extend graph convolution operation to propagate information through distant neighbors with similar topological features. We construct a dual graph with the same nodes as the original graph, but different edges representing the topological similarity of nodes. Nodes with similar topology are connected with an undirected edge. There are many ways to construct those topological edges. Here we chose to present each node as a $R^N$ vector of topological attributes and connect each node to its k most similar nodes (see Appendix~\ref{Appendix Networks Measures} for the full description of attributes used to define the topology of a node). The T-GCN includes two GCN layers performed simultaneously on the input (external features). One GCN uses the regular adjacency matrix. The second uses the dual graph. These two outputs are then concatenated to serve as an input for the next layer (typically, a standard GCN on the original graph). The network's structure is illustrated at Fig~\ref{Schematic T-GCN}.

\item Topology based GAT (T-GAT). The same as T-GCN, with GAT layers instead of GCN layers (Eq.~\ref{GAT Layer} instead of Eq.\ref{GCN Layer}).
\end{itemize}
We have also tested the  following two alternative methods to use the topology. However, both produce lower accuracies than the standard GCN.
\begin{itemize}
\item Asymmetric GCN (A-GCN). We incorporate the direction of directed networks by taking the adjacency matrix (asymmetric in directed graph) and concatenate its transpose to it – creating a $2n \times n$  matrix : $\tilde{A} = [A+I] \vert [A^T+I]$,. The dimension of the output of each layer is: $[(2N \times N)\times(N \times i_n )\times(i_n \times o_n )]=2N \times o_n$., which in turn is passed to the next layer following a rearrangement of the output by splitting and concatenating it to change dimensions from  - $2N \times O_n$  to $N \times 2O_n$. For more details about the parameters see Appendix~\ref{Appendix models parameters}. Multiple inputs were tested in these configurations as detailed in Appendix~\ref{Appendix Networks Measures}.
	
\item Combined GCN (C-GCN): This model includes two input types: a topology features matrix and an external features matrix (in the Cora and Citeseer case, the bag-of-words features). First, we pass the data matrix through a GCN layer, which leads to a $2n \times L_1$ output. The two inputs (topology and external features) are then concatenated following a rearrangement of the processed data matrix by splitting in dimension 0 and concatenating in the dimension $1 – 2n \times L_1 \rightarrow n \times 2L_1$. Following the concatenation, an $n \times (2L_1+T)$ matrix is obtained, which is passed forward to the A-GCN layers. The following layers are as above in the A-GCN. For more details about the parameters see Appendix~\ref{Appendix models parameters}. Multiple inputs were tested in these configurations as detailed in appendix~\ref{Appendix Networks Measures}.
\end{itemize}

 \section{Experimental Set-Up}
As proposed by \citet{monti2017geometric} and \citet{velivckovic2017graph}, we used one set of hyper-parameters for Cora, and used the hyper-parameters optimized for Pubmed for all other networks:
For the Cora dataset we used the following parameters. In the T-GCN, we used 1 hidden layer of size $32$ for each graph (original and dual). For the T-GAT we chose $16$ internal nodes for the regular and $8$ internal nodes for the dual graph. We also chose 8 heads (8 independent attention mechanisms, see \citet{velivckovic2017graph} for more details) for both operations at the first layer, and 1 head for the last layer (same as the original GAT).  For the other datasets, we used the optimal parameters found for PubMed: 1 hidden layer with a size of $64+16$ for T-GCN, and $16+16$ features for T-GAT. We used 16 heads on the original operation, 8 heads at the dual operation at the first layer, and 8 heads at the last layer (same as the original GAT). 

The first layer activation function was ReLU for T-GCN, and TanH for T-GAT (except for T-GAT for Cora which we used also ReLU). SoftMax was performed on the last layer of all models. Note that for T-GAT, the external features were normalized and GAT heads were concatenated on the first layer, and averaged on the last layer (same as the original GAT). See a summarized table of all parameters in Appendix \ref{Appendix models parameters}.

The input to all classification tasks was a Bag Of Words (BOW) or similar textual description. In the A-GCN and C-GCN, we included a vector of topological features (e.g. degree, clustering coefficient...). An alternative input tested is  a vector of the number of first and second neighbors belonging to each class in the training set. For example, assume a classification task with 3 possible labels, and a node with 5 neighbors and 30 second neighbors. Further assume that 1 of the first neighbors belongs to the training set and has label A, 3 of the second neighbors belong to the training set and have label A and 1 of the second neighbors belong to the training set and has label B. The input to the node would be [1,0,0,3,1,0], where the first three values represent the first neighbors and the last three values represent the second neighbors. See Appendix~\ref{Appendix Networks Measures} for more details.

\begin{figure} [t]
\begin{center}
 \includegraphics[width=14cm, height=7cm]{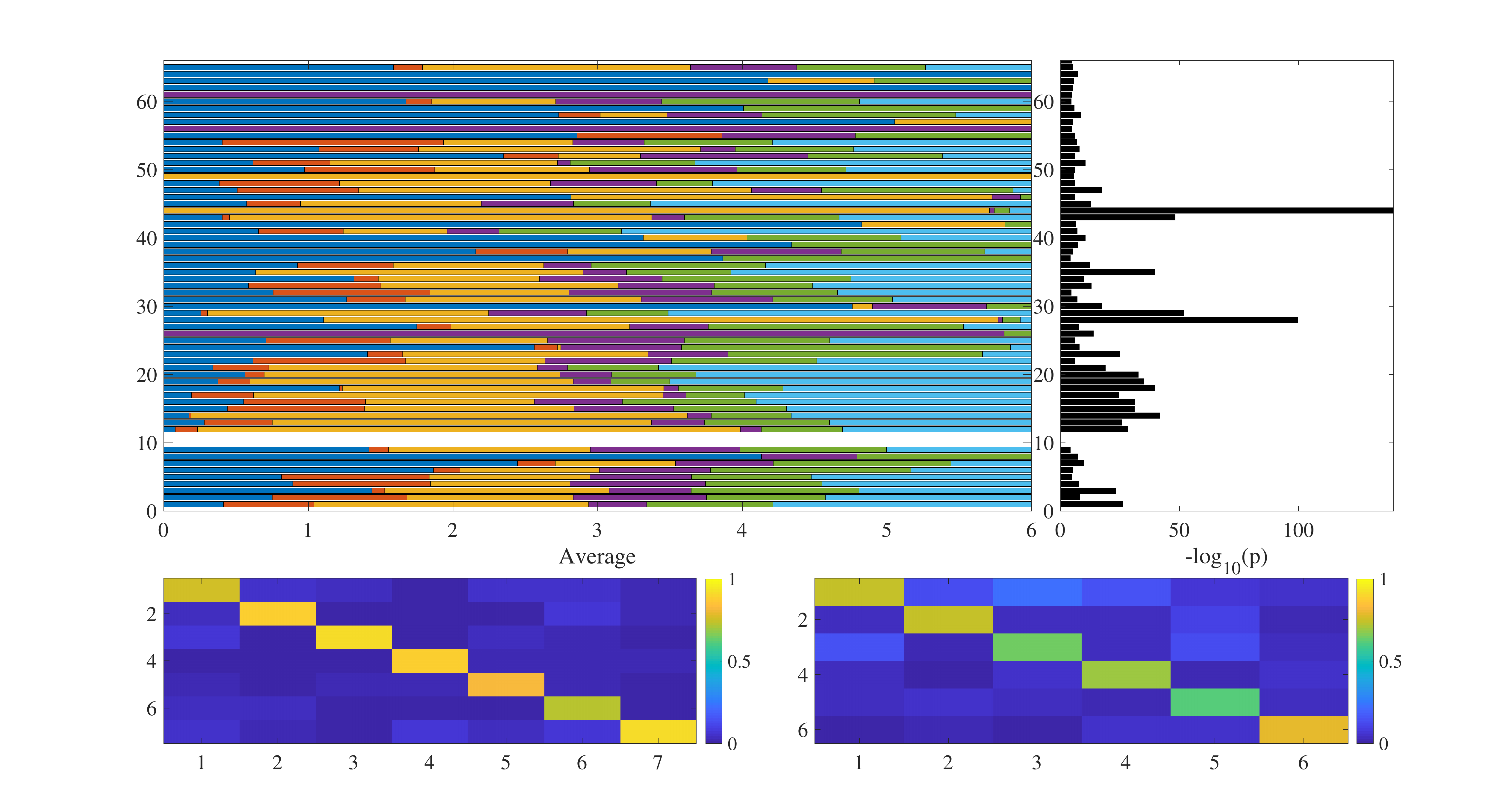}
\end{center}
\caption{\it{A) upper right plot - Log p value of non-parametric Kruskal Wallis (KW) test for the association of each topological feature with the class of the manuscripts in CiteSeer. The x axis is the feature number. One can clearly see that some features are highly associated with the manuscript class. Upper left plot Average of each topological feature for nodes belonging to a given class. All values were stacked and normalized to 1. An equal distribution would produce equally divided columns. The upper group are 4 node subgraph frequencies, and the lower group (separated by empty row) are 3 node motifs. Except for the centrality, no other node feature had a a significant KW p value after multiple measurements correction. Lower plots -  Correlation of CiteSeer and Cora manuscript class with the neighboring manuscript class. The color in each row represents the fraction of nodes neighboring a given class that belong to all possible classes. One can clearly see the dominating diagonal representing the fact that neighboring nodes tend to have the same color.}}
\label{topo correlation}
\end{figure}

\section{Results}
\subsection{topology is correlated with class}
To test that neighbor class and the node self-topology (as shall be further defined) are correlated with the node class, we performed two tests. We first computed the relative frequency of classes in neighbors, given the class of the node: 
\begin{equation}
p(neighbor \> has \> class \> i \> \vert \> current \> node\> has \> class \> j)
\end{equation}
 (Fig~\ref{topo correlation} lower plots). In the absence of correlations, one would expect a flat value, while an absolute correlation would produce an identity matrix. In the Cora or Citeseer networks, the mass of the diagonal is 60 \% of the mass (compared with an expected 15 \% ).

To test for the relation between node topology and class, we computed the average value of multiple topological features (Appendix~\ref{Appendix Networks Measures}) for nodes with a given class (in the current context manuscripts belonging to a certain field). Except for the betweenness centrality, the only topological features correlated with class were 3 and 4 small scale motif frequencies. To test for that, a Kruskal Wallis non-parametric test was performed to test for the relation between the node class (manuscript field) and the distribution of features, Over sixty different small scale motifs are  associated with the node class (Fig~\ref{topo correlation} upper plots) . 

To test that topology and information propagation can be used to classify node classes, we introduced the topological features above, and the number of neighbors belonging to the training set with a given class as input to a Feed Forward-Network (see Appendix~\ref{Appendix topo FFN}). These two types of information by themselves can be used to classify the class of nodes quite precisely (see Appendix~\ref{Appendix only topo input}).

Since the main topological factors correlated with the class are small scale motif, we propose an alternative method to test their contribution to the classification. In order to avoid the explicit computation of sub graph frequencies, which can be computationally expensive, an indirect computation of the topology can be proposed. A simple way to describe such local features is though operations on products of the adjacency matrix. For example, the number of triangles $i \rightarrow j$, $i \rightarrow k$ and $j \rightarrow k$, are the and combination of $A$ and $A*A$ (Fig~\ref{Appendix only topo input} ). Thus, instead of explicitly computing such features, one can use as input to the FFN combinations of these products on a one hot representation of the training set class. Formally, let us define for node i, the vector  $v_i$ , where $v_i^j$ is the  number of neighbors of node i that are in the training set and are of class j, and V is the matrix of all vectors $v_i$ . To this, we add a last constant value to the vector, as shall be explained. We then use different combinations of $A\times V,A^T \times V,A \times A^T \times V$ etc.  as inputs to an FFN (see Methods). 
When these products are applied to the last row (a constant value), they simply count sub-graphs. However, when multiplied by the other component, the sub-graphs composed of a specific class are counted (Fig~\ref{Appendix only topo input} B). The accuracy obtained for such products outperforms the only explicit topological measures, or information propagation (Fig~\ref{Appendix only topo input} upper plot).

\subsection{Neighbors class is a better predictor than topology in the absence of external information}
Given the correlation between the node's topology and class, we tested whether adding the topology by itself or as an additional input to the BOW would increase the prediction accuracy of the node class for the Cora and Citeseer networks. We have tested both symmetric and asymmetric GCN (see description above), and either topological input by itself or combined with the BOW. Within the topological input, we  tested three alternatives:  the number of first and second neighbors belonging to each class in the training set, the topological features of each node or their combination. 

\begin{figure} [t]
\begin{center}
 \includegraphics[width=14cm, height=7cm]{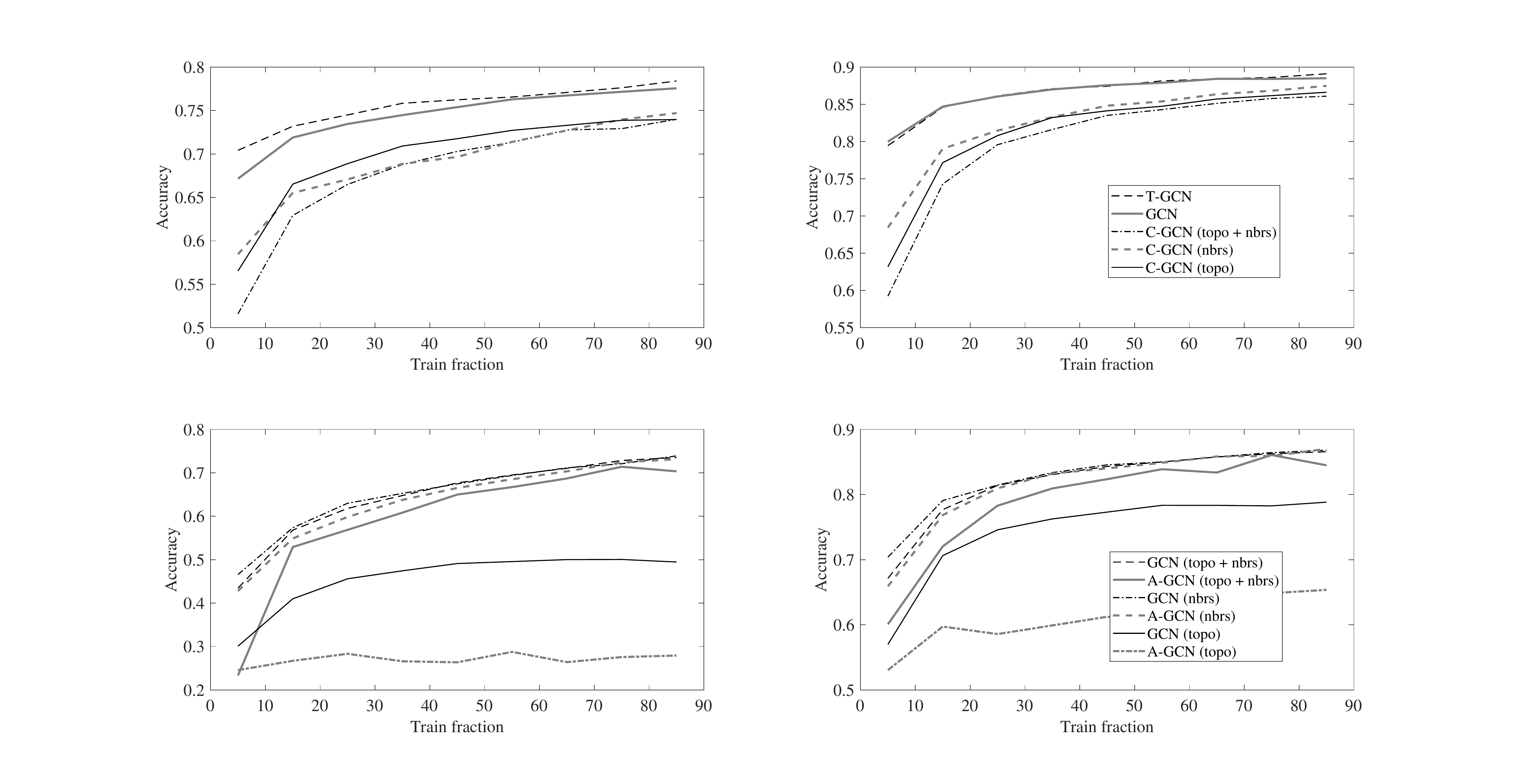}
\end{center}
\caption{\it{Average accuracy obtained with 50 random splits as a function of training size (starting from 5\%). Validation and test sets were split evenly. Upper plots are for Cora and lower plots for CiteSeer. In the right plots, we compared our asymmetric model with the standard GCN in the absence of external information. Different types of topological features were tested as input (see Results). One can clearly see that the neighbors feature is the best input (performed almost as good as external information), and the standard GCN is better than the A-GCN.
In the left plots, we compared the standard GCN with T-GCN and C-GCN (with different types of topological features), where the input is BOW. The C-GCN does not perform well with all three types, and the T-GCN is always equal to or better than the standard GCN.}}
\label{results figure}
\end{figure}

As expected, over all tested training set fractions, the models with the BOW outperform the ones without it. Within the models without BOW, ignoring the edge direction, and using only the number of neighbors in each class is better than any other combination. Moreover, combining BOW with topology as input only reduces the accuracy (Fig~\ref{results figure}). Still, it is interesting to note that the accuracy without the BOW is not so far from the accuracy with the BOW at high training set fraction (Fig~\ref{results figure}). For example, the accuracy for Cora with train size of 55\% is 87.9\% with BOW, and 85\% with only neighbors features as input.

\subsection{Model with topology with adjacency matrix outperforms all existing models}
Since adding topology as an input did not improve the accuracy, we tested whether using the topology to propagate information between distant nodes based on the similarity of topological attributes helps. We compared our topology-based convolution (T-GCN) to state of the art models on multiple standard real-world networks (see Models and Data).
In Cora, CiteSeer, and PubMed networks, we used the previously published split of train-test \citep{yang2016revisiting}, and for Cora-Full and the co-authorship networks we took 20 labeled nodes per class as the training set, 30 nodes per class as the validation set, and the rest as the test set (same as \citet{shchur2018pitfalls}). We repeated the experiment 100 times and report the average accuracy over all trials. In each trial, we split the data randomly (except for the standard splits where the train is fixed).
Parameters for all models can be found in Appendix~\ref{Appendix models parameters}.
All baselines results were copied from the related paper. Furthermore, we report the results of GCN \citep{kipf2016semi} and GAT \citep{velivckovic2017graph} using our own implementation, written using pytorch (which produce slightly lower results than the published result for the same architecture). to fairly evaluate GAT, we used 500 epochs for training. These are the base implementation used for T-GCN and T-GAT. 
The summary of the results is presented in table~\ref{results table}

\begin{table}[t]
\begin{center}
\caption{Results - average accuracy over 100 trials. For Cora, CiteSeer, and PubMed we used the standard splits as \citet{yang2016revisiting}. For Cora-Full, Physics, and CS we used $20 \times \# Classes$ random nodes as train, and $30 \times \# Classes$ for validation \citep{shchur2018pitfalls}. For Cora-Full and Physics we reported the T-GAT results of only 20 and 10 trials accordingly since they were tested on the CPU}
\vspace*{2mm}
 \label{results table}

\begin{tabular}{|r|r|r|r|r|r|r|}
\hline
    Method & {\bf CiteSeer} & {\bf Cora} & {\bf PubMed} & {\bf Physics} &   {\bf CS} & {\bf Cora Full} \\
\hline
      DCNN \citep{atwood2016diffusion} &       71.1 & 81.3 &          - &          - &          - &          - \\
\hline
 Planetoid \citep{yang2016revisiting} &         64.7 & 75.7 &       77.2 &          - &          - &          - \\
\hline
   ChebNet \citep{defferrard2016convolutional}&          69.8 &   81.2 &    74.4 &          - &          - &          - \\
\hline
       GCN \citep{kipf2016semi} &            70.3 &     81.5 &      79 &       92.8 &       91.1 & {\bf 62.2} \\
\hline
      Sage \citep{hamilton2017inductive}&            63.5 &    77.4 &     77.6 &   {\bf 93} & {\bf 91.3} &       58.6 \\
\hline
     MoNet \citep{monti2017geometric} &             - &     81.7 &      78.8 &       92.5 &       90.8 &       59.8 \\
\hline
       GAT \citep{velivckovic2017graph}&       {\bf 72.5} &    83 &         79 &       92.5 &       90.5 &       51.9 \\
\hline
    N-Sage \citep{abu2018n}&            71 &      81.8 &     79.4 &          - &          - &          - \\
\hline
     N-GCN \citep{abu2018n}&          72.2 &       83 &      79.5 &          - &          - &          - \\
\hline
 CayleyNet \citep{levie2018cayleynets}&             - &      81.9 &        - &          - &          - &          - \\
\hline
\hline
GCN (our imp) &          67.1 &     80.4 &      78.9 &      92.9 &       90.7 &       59.8 \\
\hline
GAT (our imp) &           70.9 &       83 &     78.5 &     92.2 &       89.5 &         62 \\
\hline
\hline
\bf{T-GCN} (ours) &           71.3 &   83 &   {\bf 79.8} &   {\bf 93} &       91.2 &       62.1 \\
\hline
\bf{T-GAT} (ours) &      72.2 &     {\bf 83.8} &    79.3 &          92.8 &       90.1 &          61.8 \\
\hline
\end{tabular}

\end{center}
\end{table}

One can clearly see that the T-GCN and T-GAT outperform all other models in Cora, Pubmed, and Physics (Table~\ref{results table}). Moreover, the current comparison was performed using the original split in CiteSeer, Cora, and Pubmed. We have tested random splits to check the performance of the T-GCN. Indeed, in the random split, the T-GCN always has a higher accuracy than the GCN (Fig~\ref{results figure}) even in the CiteSeer dataset, with a difference that can reach up to 3.3\%.

\section{Conclusions}
Convolution methods to aggregate information from multiple distances are among the leading image classification methods. In images, most of these convolutions are symmetric and sometimes isotropic around each point.  However, in contrast with images that are typically overlaid on a 2D lattice, graphs have a complex topology. This topology is highly informative of the properties of nodes and edges \citep{rosen2015topological, naaman2018edge}, and can thus be used to classify their classes. This complex topology can be combined with convolutional networks to improve their accuracy.

In undirected graphs, the topology can often be captured by a distance maintaining projection into $R^N$, using unsupervised methods, such as the classical MDS \citep{kruskal1964multidimensional}, or supervised methods to minimize the distance between nodes with similar classes in the training set \citep{cao2016deep}. In directed graphs, a more complex topology emerges from the asymmetry between incoming and outgoing edges (i.e., the distance between node $i$ and node $j$ differs from the distance between node $j$ and node $i$), creating a distribution of subgraphs around each node often denoted sub-graph motifs \citep{milo2002network}.  Such motifs have been reported to be associated with both single node/edge attributes as well as whole-graph attributes \citep{milo2002network}.  We have here shown that in a manuscript assignment task, the topology around each node is indeed associated with the manuscript class. 

In order to combine topological information with information propagation, we proposed a novel GCN where the fraction of second neighbors belonging to each class is used as an input, and the class of the node is compared to the softmax output of the node. This method can indeed produce a high classification accuracy, but less than the one obtained using a BOW input. Moreover, explicitly combining the topology as an input with the BOW reduces the accuracy. However, using the topology to add new edges between nodes with similar topological features actually significantly improves performance in most studied datasets. This suggests that the topology is better used to correlate between the class of distant nodes than to be actually used as an input. 

The results presented here are a combination of information propagation and topology-based classification. While each of these two elements was previously reported, their combination into a single coherent GCN based classifier provides a novel content independent method to classify nodes. With the current ever-increasing concerns about privacy, new content independent methods for node classification become essential. 

\bibliography{Topological_GCN}

\begin{thebibliography}{57}
\providecommand{\natexlab}[1]{#1}
\providecommand{\url}[1]{\texttt{#1}}
\expandafter\ifx\csname urlstyle\endcsname\relax
  \providecommand{\doi}[1]{doi: #1}\else
  \providecommand{\doi}{doi: \begingroup \urlstyle{rm}\Url}\fi

\bibitem[Abel \& Louzoun(2019)Abel and Louzoun]{abel2019regional}
Roy Abel and Yoram Louzoun.
\newblock Regional based query in graph active learning.
\newblock \emph{arXiv preprint arXiv:1906.08541}, 2019.

\bibitem[Abu-El-Haija et~al.(2018)Abu-El-Haija, Kapoor, Perozzi, and
  Lee]{abu2018n}
Sami Abu-El-Haija, Amol Kapoor, Bryan Perozzi, and Joonseok Lee.
\newblock N-gcn: Multi-scale graph convolution for semi-supervised node
  classification.
\newblock \emph{arXiv preprint arXiv:1802.08888}, 2018.

\bibitem[Atwood \& Towsley(2016)Atwood and Towsley]{atwood2016diffusion}
James Atwood and Don Towsley.
\newblock Diffusion-convolutional neural networks.
\newblock In \emph{Advances in Neural Information Processing Systems}, pp.\
  1993--2001, 2016.

\bibitem[Batagelj \& Zaversnik(2003)Batagelj and Zaversnik]{batagelj2003m}
Vladimir Batagelj and Matjaz Zaversnik.
\newblock An o (m) algorithm for cores decomposition of networks.
\newblock \emph{arXiv preprint cs/0310049}, 2003.

\bibitem[Belkin \& Niyogi(2002)Belkin and Niyogi]{belkin2002laplacian}
Mikhail Belkin and Partha Niyogi.
\newblock Laplacian eigenmaps and spectral techniques for embedding and
  clustering.
\newblock In \emph{Advances in neural information processing systems}, pp.\
  585--591, 2002.

\bibitem[Belkin \& Niyogi(2004)Belkin and Niyogi]{belkin2004semi}
Mikhail Belkin and Partha Niyogi.
\newblock Semi-supervised learning on riemannian manifolds.
\newblock \emph{Machine learning}, 56\penalty0 (1-3):\penalty0 209--239, 2004.

\bibitem[Berberidis \& Giannakis(2018)Berberidis and
  Giannakis]{berberidis2018data}
Dimitris Berberidis and Georgios~B Giannakis.
\newblock Data-adaptive active sampling for efficient graph-cognizant
  classification.
\newblock \emph{IEEE Transactions on Signal Processing}, 66\penalty0
  (19):\penalty0 5167--5179, 2018.

\bibitem[Blondel et~al.(2008)Blondel, Guillaume, Lambiotte, and
  Lefebvre]{blondel2008fast}
Vincent~D Blondel, Jean-Loup Guillaume, Renaud Lambiotte, and Etienne Lefebvre.
\newblock Fast unfolding of communities in large networks.
\newblock \emph{Journal of statistical mechanics: theory and experiment},
  2008\penalty0 (10):\penalty0 P10008, 2008.

\bibitem[Bojchevski \& G{\"u}nnemann(2017)Bojchevski and
  G{\"u}nnemann]{bojchevski2017deep}
Aleksandar Bojchevski and Stephan G{\"u}nnemann.
\newblock Deep gaussian embedding of graphs: Unsupervised inductive learning
  via ranking.
\newblock \emph{arXiv preprint arXiv:1707.03815}, 2017.

\bibitem[Bronstein et~al.(2017)Bronstein, Bruna, LeCun, Szlam, and
  Vandergheynst]{bronstein2017geometric}
Michael~M Bronstein, Joan Bruna, Yann LeCun, Arthur Szlam, and Pierre
  Vandergheynst.
\newblock Geometric deep learning: going beyond euclidean data.
\newblock \emph{IEEE Signal Processing Magazine}, 34\penalty0 (4):\penalty0
  18--42, 2017.

\bibitem[Bruna et~al.(2013)Bruna, Zaremba, Szlam, and LeCun]{bruna2013spectral}
Joan Bruna, Wojciech Zaremba, Arthur Szlam, and Yann LeCun.
\newblock Spectral networks and locally connected networks on graphs.
\newblock \emph{arXiv preprint arXiv:1312.6203}, 2013.

\bibitem[Cannistraci et~al.(2013)Cannistraci, Alanis-Lobato, and
  Ravasi]{cannistraci2013minimum}
Carlo~Vittorio Cannistraci, Gregorio Alanis-Lobato, and Timothy Ravasi.
\newblock Minimum curvilinearity to enhance topological prediction of protein
  interactions by network embedding.
\newblock \emph{Bioinformatics}, 29\penalty0 (13):\penalty0 i199--i209, 2013.

\bibitem[Cao et~al.(2016)Cao, Lu, and Xu]{cao2016deep}
Shaosheng Cao, Wei Lu, and Qiongkai Xu.
\newblock Deep neural networks for learning graph representations.
\newblock In \emph{Thirtieth AAAI Conference on Artificial Intelligence}, 2016.

\bibitem[Defferrard et~al.(2016)Defferrard, Bresson, and
  Vandergheynst]{defferrard2016convolutional}
Micha{\"e}l Defferrard, Xavier Bresson, and Pierre Vandergheynst.
\newblock Convolutional neural networks on graphs with fast localized spectral
  filtering.
\newblock In \emph{Advances in neural information processing systems}, pp.\
  3844--3852, 2016.

\bibitem[Dhillon et~al.(2007)Dhillon, Guan, and Kulis]{dhillon2007weighted}
Inderjit~S Dhillon, Yuqiang Guan, and Brian Kulis.
\newblock Weighted graph cuts without eigenvectors a multilevel approach.
\newblock \emph{IEEE transactions on pattern analysis and machine
  intelligence}, 29\penalty0 (11):\penalty0 1944--1957, 2007.

\bibitem[Dijkstra(1959)]{dijkstra1959note}
Edsger~W Dijkstra.
\newblock A note on two problems in connexion with graphs.
\newblock \emph{Numerische mathematik}, 1\penalty0 (1):\penalty0 269--271,
  1959.

\bibitem[Duvenaud et~al.(2015)Duvenaud, Maclaurin, Iparraguirre, Bombarell,
  Hirzel, Aspuru-Guzik, and Adams]{duvenaud2015convolutional}
David~K Duvenaud, Dougal Maclaurin, Jorge Iparraguirre, Rafael Bombarell,
  Timothy Hirzel, Al{\'a}n Aspuru-Guzik, and Ryan~P Adams.
\newblock Convolutional networks on graphs for learning molecular fingerprints.
\newblock In \emph{Advances in neural information processing systems}, pp.\
  2224--2232, 2015.

\bibitem[Everett \& Borgatti(1999)Everett and Borgatti]{everett1999centrality}
Martin~G Everett and Stephen~P Borgatti.
\newblock The centrality of groups and classes.
\newblock \emph{The Journal of mathematical sociology}, 23\penalty0
  (3):\penalty0 181--201, 1999.

\bibitem[Fey et~al.(2018)Fey, Eric~Lenssen, Weichert, and
  M{\"u}ller]{fey2018splinecnn}
Matthias Fey, Jan Eric~Lenssen, Frank Weichert, and Heinrich M{\"u}ller.
\newblock Splinecnn: Fast geometric deep learning with continuous b-spline
  kernels.
\newblock In \emph{Proceedings of the IEEE Conference on Computer Vision and
  Pattern Recognition}, pp.\  869--877, 2018.

\bibitem[Gori et~al.(2005)Gori, Monfardini, and Scarselli]{gori2005new}
Marco Gori, Gabriele Monfardini, and Franco Scarselli.
\newblock A new model for learning in graph domains.
\newblock In \emph{Proceedings. 2005 IEEE International Joint Conference on
  Neural Networks, 2005.}, volume~2, pp.\  729--734. IEEE, 2005.

\bibitem[Grover \& Leskovec(2016)Grover and Leskovec]{grover2016node2vec}
Aditya Grover and Jure Leskovec.
\newblock node2vec: Scalable feature learning for networks.
\newblock In \emph{Proceedings of the 22nd ACM SIGKDD international conference
  on Knowledge discovery and data mining}, pp.\  855--864. ACM, 2016.

\bibitem[Hamilton et~al.(2017)Hamilton, Ying, and
  Leskovec]{hamilton2017inductive}
Will Hamilton, Zhitao Ying, and Jure Leskovec.
\newblock Inductive representation learning on large graphs.
\newblock In \emph{Advances in Neural Information Processing Systems}, pp.\
  1024--1034, 2017.

\bibitem[Henaff et~al.(2015)Henaff, Bruna, and LeCun]{henaff2015deep}
Mikael Henaff, Joan Bruna, and Yann LeCun.
\newblock Deep convolutional networks on graph-structured data.
\newblock \emph{arXiv preprint arXiv:1506.05163}, 2015.

\bibitem[Henderson et~al.(2011)Henderson, Gallagher, Li, Akoglu, Eliassi-Rad,
  Tong, and Faloutsos]{henderson2011s}
Keith Henderson, Brian Gallagher, Lei Li, Leman Akoglu, Tina Eliassi-Rad,
  Hanghang Tong, and Christos Faloutsos.
\newblock It's who you know: graph mining using recursive structural features.
\newblock In \emph{Proceedings of the 17th ACM SIGKDD international conference
  on Knowledge discovery and data mining}, pp.\  663--671. ACM, 2011.

\bibitem[Itzhack et~al.(2007)Itzhack, Mogilevski, and
  Louzoun]{itzhack2007optimal}
Royi Itzhack, Yelena Mogilevski, and Yoram Louzoun.
\newblock An optimal algorithm for counting network motifs.
\newblock \emph{Physica A: Statistical Mechanics and its Applications},
  381:\penalty0 482--490, 2007.

\bibitem[Ji \& Han(2012)Ji and Han]{ji2012variance}
Ming Ji and Jiawei Han.
\newblock A variance minimization criterion to active learning on graphs.
\newblock In \emph{Artificial Intelligence and Statistics}, pp.\  556--564,
  2012.

\bibitem[Karypis \& Kumar(1995)Karypis and Kumar]{karypis1995metis}
George Karypis and Vipin Kumar.
\newblock Metis--unstructured graph partitioning and sparse matrix ordering
  system, version 2.0.
\newblock 1995.

\bibitem[Kipf \& Welling(2016)Kipf and Welling]{kipf2016semi}
Thomas~N Kipf and Max Welling.
\newblock Semi-supervised classification with graph convolutional networks.
\newblock \emph{arXiv preprint arXiv:1609.02907}, 2016.

\bibitem[Kruskal(1964)]{kruskal1964multidimensional}
Joseph~B Kruskal.
\newblock Multidimensional scaling by optimizing goodness of fit to a nonmetric
  hypothesis.
\newblock \emph{Psychometrika}, 29\penalty0 (1):\penalty0 1--27, 1964.

\bibitem[Lei et~al.(2019)Lei, Qin, Bai, Zhang, and Yang]{lei2019gcn}
Kai Lei, Meng Qin, Bo~Bai, Gong Zhang, and Min Yang.
\newblock Gcn-gan: A non-linear temporal link prediction model for weighted
  dynamic networks.
\newblock In \emph{IEEE INFOCOM 2019-IEEE Conference on Computer
  Communications}, pp.\  388--396. IEEE, 2019.

\bibitem[Levie et~al.(2018)Levie, Monti, Bresson, and
  Bronstein]{levie2018cayleynets}
Ron Levie, Federico Monti, Xavier Bresson, and Michael~M Bronstein.
\newblock Cayleynets: Graph convolutional neural networks with complex rational
  spectral filters.
\newblock \emph{IEEE Transactions on Signal Processing}, 67\penalty0
  (1):\penalty0 97--109, 2018.

\bibitem[Levy et~al.(2015)Levy, Goldberg, and Dagan]{levy2015improving}
Omer Levy, Yoav Goldberg, and Ido Dagan.
\newblock Improving distributional similarity with lessons learned from word
  embeddings.
\newblock \emph{Transactions of the Association for Computational Linguistics},
  3:\penalty0 211--225, 2015.

\bibitem[Li et~al.(2015)Li, Tarlow, Brockschmidt, and Zemel]{li2015gated}
Yujia Li, Daniel Tarlow, Marc Brockschmidt, and Richard Zemel.
\newblock Gated graph sequence neural networks.
\newblock \emph{arXiv preprint arXiv:1511.05493}, 2015.

\bibitem[Ling et~al.(2019)Ling, Gao, Kar, Chen, and Fidler]{ling2019fast}
Huan Ling, Jun Gao, Amlan Kar, Wenzheng Chen, and Sanja Fidler.
\newblock Fast interactive object annotation with curve-gcn.
\newblock In \emph{Proceedings of the IEEE Conference on Computer Vision and
  Pattern Recognition}, pp.\  5257--5266, 2019.

\bibitem[Masci et~al.(2015)Masci, Boscaini, Bronstein, and
  Vandergheynst]{masci2015shapenet}
Jonathan Masci, Davide Boscaini, Michael Bronstein, and Pierre Vandergheynst.
\newblock Shapenet: Convolutional neural networks on non-euclidean manifolds.
\newblock Technical report, 2015.

\bibitem[Mikolov et~al.(2013)Mikolov, Sutskever, Chen, Corrado, and
  Dean]{mikolov2013distributed}
Tomas Mikolov, Ilya Sutskever, Kai Chen, Greg~S Corrado, and Jeff Dean.
\newblock Distributed representations of words and phrases and their
  compositionality.
\newblock In \emph{Advances in neural information processing systems}, pp.\
  3111--3119, 2013.

\bibitem[Milo et~al.(2002)Milo, Shen-Orr, Itzkovitz, Kashtan, Chklovskii, and
  Alon]{milo2002network}
Ron Milo, Shai Shen-Orr, Shalev Itzkovitz, Nadav Kashtan, Dmitri Chklovskii,
  and Uri Alon.
\newblock Network motifs: simple building blocks of complex networks.
\newblock \emph{Science}, 298\penalty0 (5594):\penalty0 824--827, 2002.

\bibitem[Monti et~al.(2017)Monti, Boscaini, Masci, Rodola, Svoboda, and
  Bronstein]{monti2017geometric}
Federico Monti, Davide Boscaini, Jonathan Masci, Emanuele Rodola, Jan Svoboda,
  and Michael~M Bronstein.
\newblock Geometric deep learning on graphs and manifolds using mixture model
  cnns.
\newblock In \emph{Proceedings of the IEEE Conference on Computer Vision and
  Pattern Recognition}, pp.\  5115--5124, 2017.

\bibitem[Muchnik et~al.(2007)Muchnik, Itzhack, Solomon, and
  Louzoun]{muchnik2007self}
Lev Muchnik, Royi Itzhack, Sorin Solomon, and Yoram Louzoun.
\newblock Self-emergence of knowledge trees: Extraction of the wikipedia
  hierarchies.
\newblock \emph{Physical Review E}, 76\penalty0 (1):\penalty0 016106, 2007.

\bibitem[Naaman et~al.(2018)Naaman, Cohen, and Louzoun]{naaman2018edge}
Roi Naaman, Keren Cohen, and Yoram Louzoun.
\newblock Edge sign prediction based on a combination of network structural
  topology and sign propagation.
\newblock \emph{Journal of Complex Networks}, 7\penalty0 (1):\penalty0 54--66,
  2018.

\bibitem[Perozzi et~al.(2014)Perozzi, Al-Rfou, and Skiena]{perozzi2014deepwalk}
Bryan Perozzi, Rami Al-Rfou, and Steven Skiena.
\newblock Deepwalk: Online learning of social representations.
\newblock In \emph{Proceedings of the 20th ACM SIGKDD international conference
  on Knowledge discovery and data mining}, pp.\  701--710. ACM, 2014.

\bibitem[Rosen \& Louzoun(2014)Rosen and Louzoun]{rosen2014directionality}
Yonatan Rosen and Yoram Louzoun.
\newblock Directionality of real world networks as predicted by path length in
  directed and undirected graphs.
\newblock \emph{Physica A: Statistical Mechanics and Its Applications},
  401:\penalty0 118--129, 2014.

\bibitem[Rosen \& Louzoun(2015)Rosen and Louzoun]{rosen2015topological}
Yonatan Rosen and Yoram Louzoun.
\newblock Topological similarity as a proxy to content similarity.
\newblock \emph{Journal of Complex Networks}, 4\penalty0 (1):\penalty0 38--60,
  2015.

\bibitem[Rosenfeld \& Globerson(2017)Rosenfeld and
  Globerson]{rosenfeld2017semi}
Nir Rosenfeld and Amir Globerson.
\newblock Semi-supervised learning with competitive infection models.
\newblock \emph{arXiv preprint arXiv:1703.06426}, 2017.

\bibitem[Scarselli et~al.(2008)Scarselli, Gori, Tsoi, Hagenbuchner, and
  Monfardini]{scarselli2008graph}
Franco Scarselli, Marco Gori, Ah~Chung Tsoi, Markus Hagenbuchner, and Gabriele
  Monfardini.
\newblock The graph neural network model.
\newblock \emph{IEEE Transactions on Neural Networks}, 20\penalty0
  (1):\penalty0 61--80, 2008.

\bibitem[Schlichtkrull et~al.(2018)Schlichtkrull, Kipf, Bloem, Van Den~Berg,
  Titov, and Welling]{schlichtkrull2018modeling}
Michael Schlichtkrull, Thomas~N Kipf, Peter Bloem, Rianne Van Den~Berg, Ivan
  Titov, and Max Welling.
\newblock Modeling relational data with graph convolutional networks.
\newblock In \emph{European Semantic Web Conference}, pp.\  593--607. Springer,
  2018.

\bibitem[Seo et~al.(2018)Seo, Defferrard, Vandergheynst, and
  Bresson]{seo2018structured}
Youngjoo Seo, Micha{\"e}l Defferrard, Pierre Vandergheynst, and Xavier Bresson.
\newblock Structured sequence modeling with graph convolutional recurrent
  networks.
\newblock In \emph{International Conference on Neural Information Processing},
  pp.\  362--373. Springer, 2018.

\bibitem[Shchur et~al.(2018)Shchur, Mumme, Bojchevski, and
  G{\"u}nnemann]{shchur2018pitfalls}
Oleksandr Shchur, Maximilian Mumme, Aleksandar Bojchevski, and Stephan
  G{\"u}nnemann.
\newblock Pitfalls of graph neural network evaluation.
\newblock \emph{arXiv preprint arXiv:1811.05868}, 2018.

\bibitem[Shi \& Malik(2000)Shi and Malik]{shi2000normalized}
Jianbo Shi and Jitendra Malik.
\newblock Normalized cuts and image segmentation.
\newblock \emph{Departmental Papers (CIS)}, pp.\  107, 2000.

\bibitem[Sindhwani et~al.(2005)Sindhwani, Niyogi, and
  Belkin]{sindhwani2005beyond}
Vikas Sindhwani, Partha Niyogi, and Mikhail Belkin.
\newblock Beyond the point cloud: from transductive to semi-supervised
  learning.
\newblock In \emph{Proceedings of the 22nd international conference on Machine
  learning}, pp.\  824--831. ACM, 2005.

\bibitem[Veli{\v{c}}kovi{\'c} et~al.(2017)Veli{\v{c}}kovi{\'c}, Cucurull,
  Casanova, Romero, Lio, and Bengio]{velivckovic2017graph}
Petar Veli{\v{c}}kovi{\'c}, Guillem Cucurull, Arantxa Casanova, Adriana Romero,
  Pietro Lio, and Yoshua Bengio.
\newblock Graph attention networks.
\newblock \emph{arXiv preprint arXiv:1710.10903}, 2017.

\bibitem[Wang et~al.(2018)Wang, Girshick, Gupta, and He]{wang2018non}
Xiaolong Wang, Ross Girshick, Abhinav Gupta, and Kaiming He.
\newblock Non-local neural networks.
\newblock In \emph{Proceedings of the IEEE Conference on Computer Vision and
  Pattern Recognition}, pp.\  7794--7803, 2018.

\bibitem[Yang et~al.(2013)Yang, McAuley, and Leskovec]{yang2013community}
Jaewon Yang, Julian McAuley, and Jure Leskovec.
\newblock Community detection in networks with node attributes.
\newblock In \emph{2013 IEEE 13th International Conference on Data Mining},
  pp.\  1151--1156. IEEE, 2013.

\bibitem[Yang et~al.(2016)Yang, Cohen, and Salakhutdinov]{yang2016revisiting}
Zhilin Yang, William~W Cohen, and Ruslan Salakhutdinov.
\newblock Revisiting semi-supervised learning with graph embeddings.
\newblock \emph{arXiv preprint arXiv:1603.08861}, 2016.

\bibitem[Zhou et~al.(2004)Zhou, Bousquet, Lal, Weston, and
  Sch{\"o}lkopf]{zhou2004learning}
Dengyong Zhou, Olivier Bousquet, Thomas~N Lal, Jason Weston, and Bernhard
  Sch{\"o}lkopf.
\newblock Learning with local and global consistency.
\newblock In \emph{Advances in neural information processing systems}, pp.\
  321--328, 2004.

\bibitem[Zhu et~al.(2003{\natexlab{a}})Zhu, Ghahramani, and
  Lafferty]{zhu2003semi}
Xiaojin Zhu, Zoubin Ghahramani, and John~D Lafferty.
\newblock Semi-supervised learning using gaussian fields and harmonic
  functions.
\newblock In \emph{Proceedings of the 20th International conference on Machine
  learning (ICML-03)}, pp.\  912--919, 2003{\natexlab{a}}.

\bibitem[Zhu et~al.(2003{\natexlab{b}})Zhu, Lafferty, and
  Ghahramani]{zhu2003combining}
Xiaojin Zhu, John Lafferty, and Zoubin Ghahramani.
\newblock Combining active learning and semi-supervised learning using gaussian
  fields and harmonic functions.
\newblock In \emph{ICML 2003 workshop on the continuum from labeled to
  unlabeled data in machine learning and data mining}, volume~3,
  2003{\natexlab{b}}.

\end{thebibliography}
\bibliographystyle{iclr2020_conference}

\appendix
\section{Appendix}
\subsection{Datasets studied} \label{Appendix data sets}
The citation networks contain scientific papers divided into classes by their research field. Edges describe citations in the data set. BOW is also available to describe each publication in the dataset. BOW can be either a $1$/$0$ vector or a TF/IDF weighted word vector for PubMed.

Coauthor CS and Coauthor Physics are co-authorship graphs based on the Microsoft Academic Graph from the KDD Cup 2016 challenge 3. Here, nodes are authors, that are connected by an edge if they co-authored a paper, node features represent paper keywords for each author’s papers, and class labels indicate the most active fields of study for each author.

\begin{table}[htbp]
\begin{center}
\caption{Datasets statistics}
\vspace*{2mm}
\begin{tabular}{|c|c|c|c|c|}
\hline
\textbf{Model} & \textbf{\textit{Nodes}}& \textbf{\textit{Edges}}& \textbf{\textit{Classes}} & \textbf{\textit{Features}}\\
\hline
CORA&	2,708&	5,429&	7& 1433\\
\hline
CITESEER&	3,312&	4,732&	6& 3703 \\
\hline
PubMed&	19,717&	44,324&	3& 500\\
\hline
Cora-Full&	18,703&	62,421&	67& 8710 \\
\hline
Co-Author CS&	18,333&	81,894&	15& 6805\\
\hline
Co-Author Physics&	34,493&	247,962&	5 & 8415\\
\hline

\end{tabular}
\label{data table}
\end{center}
\end{table}

\subsection{Models Parameters} \label{Appendix models parameters}
Here are the parameters used for each of the models. For T-GCN and T-GAT the parameters were optimized for PubMed ( as observed by \citet{monti2017geometric} and \citet{velivckovic2017graph}) except for Cora data for set which we used slightly different parameters (denotes as T-GCN Cora, and T-GAT Cora).
The parameters are summarized in Table~\ref{Model Parameters}.

In all models, the activation function of the last layer is Softmax. The activation function of the first layer is presented In Table~\ref{Model Parameters}.
Hidden size X+Y means size of X for the original GCN operator and Y for the GCN on the dual graph. The two outputs are concatenated to a total of X+Y size. 
GAT heads X,Y,Z means X heads for the original GAT operator, and Y heads for the GAT on the dual graph. Z is the number of heads in the last layer. See Models And Data for more details.

\begin{table}[htbp]
\begin{center}
\caption{Models Parameters}
\vspace*{2mm}
\begin{tabular}{|c|c|c|c|c|}
\hline
\textbf{Model} & \textbf{T-GCN}& \textbf{T-GCN (CORA)}& \textbf{T-GAT}& \textbf{T-GAT (CORA)} \\
\hline
\textbf{Activation}&	TanH&	ReLU&	ReLU& ReLU\\
\hline
\textbf{Drop Out}&	0.7&	0.6&	0.6& 0.7 \\
\hline
\textbf{Hidden Size}&	64+16&	32+32&	16+16& 16+8\\
\hline
\textbf{Learning Rate}&	0.01&	0.001&	0.01& 0.01 \\
\hline
\textbf{Weight Decay}&	0.0005&	0.01&	0.0005& 0.001\\
\hline
\textbf{Epochs}&	400&	300&	400 & 500\\
\hline
\textbf{K-NN}&	8&	8&	15 & 10\\
\hline
\textbf{Normalized Features}&	True&	False&	True & True\\
\hline
\textbf{GAT Heads}&	-&	-&	16,8,8 & 8,8,1\\
\hline

\end{tabular}

\vspace*{2mm}
\begin{tabular}{|c|c|c|}
\hline
\textbf{Model} & \textbf{A-GCN}& \textbf{C-GCN} \\
\hline
\textbf{Activation}&	ReLU&	ReLU\\
\hline
\textbf{Drop Out}&	0.6&	0.6 \\
\hline
\textbf{Hidden Size}&	100,35&	16\\
\hline
\textbf{Learning Rate}&	0.01&	0.01 \\
\hline
\textbf{Weight Decay}&	0.001&	0.001\\
\hline
\textbf{Epochs}&	200&	200\\
\hline

\end{tabular}
\label{Model Parameters}
\end{center}
\end{table}

\subsection{Networks measures} \label{Appendix Networks Measures}
Our goal is to use the graph structure to classify node colors. Hence, we compute features that are only based on the graph structure, ignoring any external content associated with each node. Those features are used to convert nodes into the appropriate network attribute vector (NAV) \citep{naaman2018edge}. Following is a list of attributes used. Note that other attributes may have been used with probably similar results. 
\begin{itemize}
\item Degree -number of in and out (in case of directed graphs) edges. 
\item Betweenness Centrality \cite{everett1999centrality}. Betweenness is a centrality measure of a vertex. It is defined by the numbers of shortest paths from all vertices that pass through the vertex.
\item	Closeness Centrality. Closeness is a centrality measure of a vertex. It is defined as the average length of the shortest path between the vertex and all other vertices in the graph.
\item	Distance distribution. We compute the distribution of distances from each node to all other nodes using a Djekstra algorithm \cite{dijkstra1959note}, and then use the first and second moments of this distribution. 
\item	Flow \citep{rosen2014directionality}. We define the flow measure of a node as the ratio between the undirected and directed distances between the node and all other nodes. 
\item	Attraction \citep{muchnik2007self} . Attraction Basin hierarchy is the comparison between the weighted fraction of the network that can be reached from each vertex with the weighted fraction of the network from which the vertex can be reached.
\item	Motifs Network motifs are small connected sub-graphs. We use an extension of the \citet{itzhack2007optimal}  algorithm to calculate motifs. For each node, we compute the frequency of each motif where this node participates.  
\item	 K-cores \citep{batagelj2003m}.  A K-core is a maximal subgraph that contains vertices of degree k or more. Equivalently, it is the subgraph of G formed by repeatedly deleting all nodes of degree less than k.
\item	 Louvain community detection algorithm \cite{blondel2008fast}. The Louvain algorithm is a community detection algorithm. The algorithm works by optimization of modularity, which has a scale value between -1 to 1.
\end{itemize}

Neighbors Feature. We also used a feature of the training set labels. We summed for each node the number of neighbors belonging to each class in the training set. The sum was represented as a vector of sums (e.g. if a node has 10 neighbors, only three of which are in the training set, with two belonging to the first class, and one belonging to the third class, the vector would be $[2,0,1,..]$). The sum was performed on first and second neighbors producing a vector of twice the number of classes. In a directed graph we calculated two features, one for In neighbors and the second for Out neighbors.

\subsection{Feed Forward Network} \label{Appendix topo FFN}
The results in Figure 2 are produced through a feed-forward network with two internal layers of sizes 300 and 100 internal nodes and an output layer with the number of possible classifications (7 and 6 in CiteSeer and Cora, respectively). The nonlinearities were Relu’s in the internal layers and a linear function in the output layer. An L2 regularization of 0.2 was used for all layers and a $10\%$ drop in our rate.  The loss function was a categorical cross-entropy as implemented in Keras with a TensorFlow backend.

\begin{figure} [t]
\begin{center}
 \includegraphics[width=12cm, height=15cm]{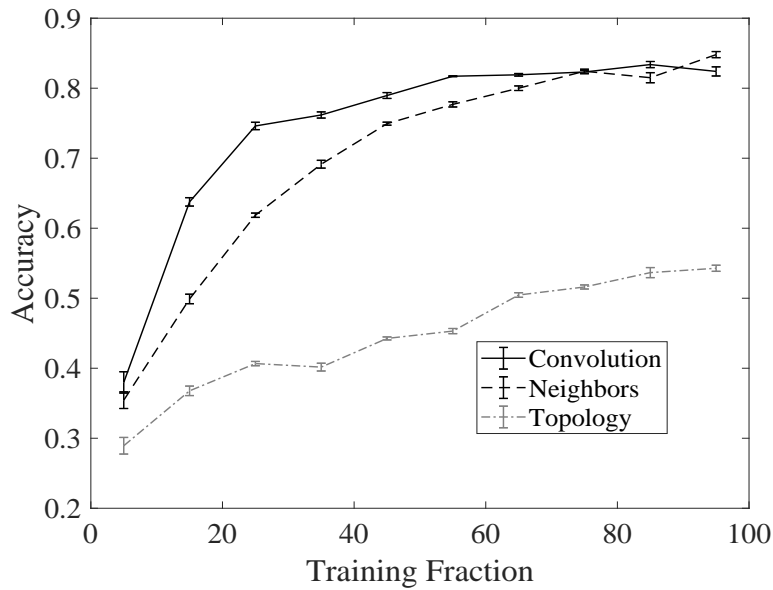} 
 \includegraphics[width=7cm, height=7cm]{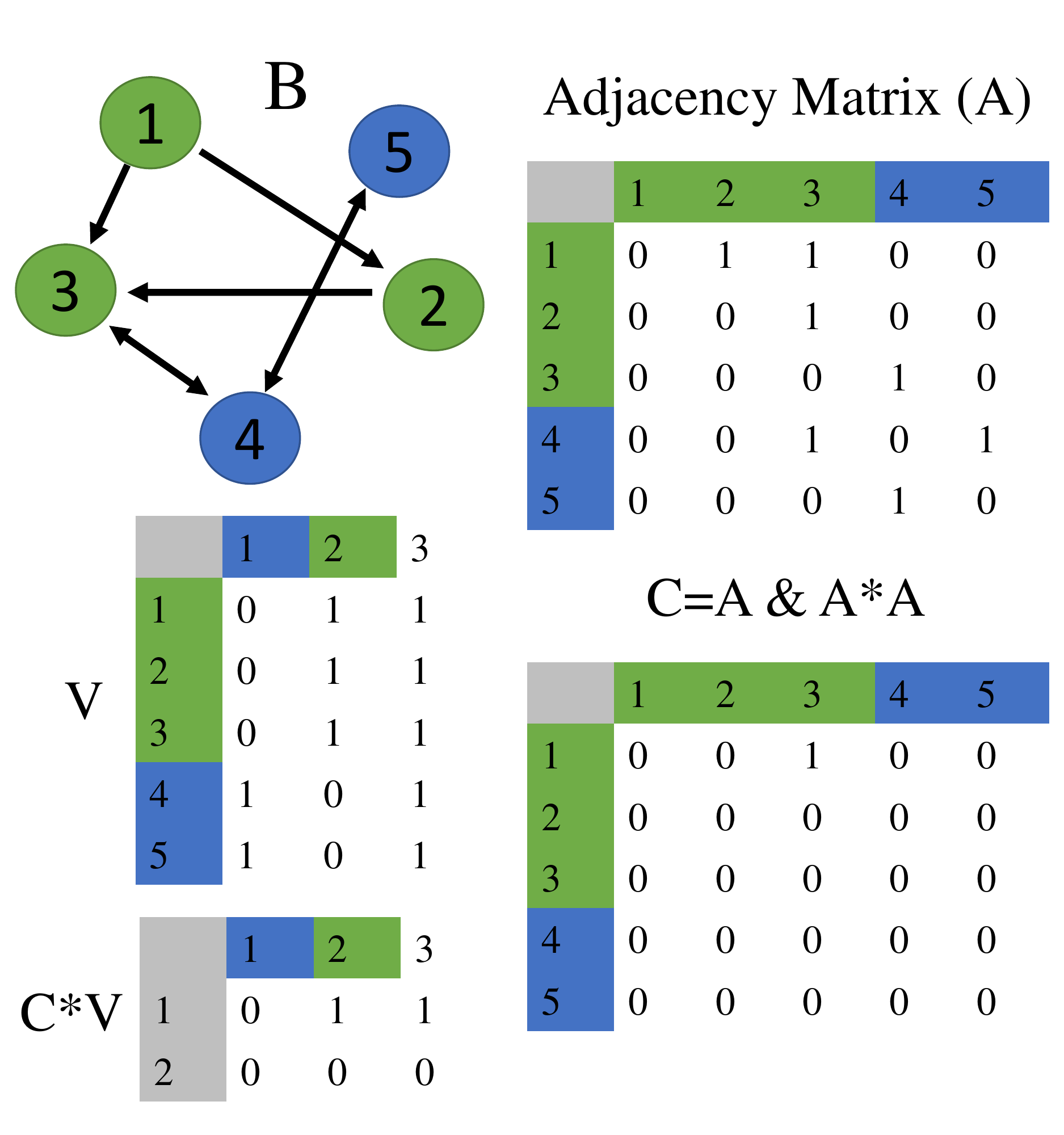}
\end{center}
\caption{\it{Upper plot -- Accuracy of the CiteSeer classification when only topological features are used (dotted dashed line), when each node is classified using the distribution of classes in its neighbors (dashed), or when products of the neighbors and different combinations of the adjacency matrix are used (full line). Lower plot.  Example of sub-graph frequency through adjacency matrix products.  Given the graph plotted on the left, the appropriate adjacency matrix (A) and a division of the nodes into green and blue nodes, one can count the number of feed-forward motifs ($x \rightarrow y$ and $x \rightarrow z \rightarrow y$) originating from green and blue nodes through the product of $C=A AND A*A$ with the color one-hot matrix of the nodes (V). One can see that there is a total of 1 such triangle and it originates from a green node. The last column in $V$ and $C*V$ is simply the sum of the row.}}

\label{Appendix only topo input}
\end{figure}

\end{document}